\documentclass[conference,letterpaper]{IEEEtran}
\usepackage{amsmath,amsfonts,amssymb}
\usepackage{algorithmic}
\usepackage{algorithm}
\usepackage{amsthm}
\usepackage{array}
\usepackage[caption=false]{subfig}
\usepackage{textcomp}
\usepackage{stfloats}
\usepackage{mathtools, cuted}
\usepackage{url}
\usepackage{booktabs}
\DeclareMathOperator{\diag}{diag}
\usepackage{bm}
\usepackage[utf8]{inputenc}
\usepackage[T1]{fontenc}
\usepackage{hyperref}
\usepackage{verbatim}
\usepackage{graphicx}
\usepackage[noadjust]{cite}
\usepackage[thinc]{esdiff}
\usepackage{xcolor}
\newtheorem{theorem}{Theorem}

\newcommand{\bmone}{\bm 1}
\newcommand{\bmu}{\bm\mu}
\newcommand{\bnu}{\bm\nu}
\newcommand{\Cov}{\operatorname{Cov}}
\newcommand{\Var}{\operatorname{Var}}

\ifCLASSINFOpdf
\else
\fi

\begin{document}
%
\title{Joint Freshness and Age-Dispersion Control over Finite-State Markov Wireless Channels}
%
%
%


\author{
    \IEEEauthorblockN{
	Aresh Dadlani\IEEEauthorrefmark{1}, 
    Hina Tabassum\IEEEauthorrefmark{2}, 
	Muthukrishnan Senthil Kumar\IEEEauthorrefmark{3}, and 
    Masoumeh Moradian\IEEEauthorrefmark{4}
    }
    \IEEEauthorblockA{
        \IEEEauthorrefmark{1}School of Computing Sciences and Mathematics, Mount Royal University, Calgary, Canada\\
        \IEEEauthorrefmark{2}Department of Electrical Engineering and Computer Science, York University, Toronto, Canada\\
	    \IEEEauthorrefmark{3}Department of Applied Mathematics and Computational Sciences, PSG College of Technology, Coimbatore, India\\
        \IEEEauthorrefmark{4}School of Computer Engineering, K. N. Toosi University of Technology, Tehran, Iran\\
	Emails: adadlani@mtroyal.ca, hinat@yorku.ca, msk.amcs@psgtech.ac.in, mmoradian@kntu.ac.ir
    }
}


%
%

\maketitle

\begin{abstract}
\fontdimen2\font=0.7ex
Age of information (AoI) has become a standard design objective for timely monitoring as it measures the freshness of the latest update at a receiver. AoI alone, however, is insufficient in goal-oriented applications where decisions depend on consecutive observations. Age dispersion complements AoI by measuring the generation-time separation between consecutive updates. In this paper, we study joint freshness and dispersion control for a generate-at-will status update link modeled as a finite-state Markov wireless channel. The objective is to minimize the long-run probability that either AoI or age dispersion exceeds a prescribed threshold under an average transmission~rate constraint. For channel-dependent randomized transmission policies, we derive exact matrix expressions for the stationary joint~distribution of AoI and dispersion. For reversible channels, mean dispersion is lower bounded by the reciprocal of the delivery throughput, and the difference is an explicit non-negative variance term for which we establish the exact equality condition. We then formulate the adaptive joint-threshold control problem as a constrained Markov decision process and prove that it admits an exact finite-state representation whose size scales linearly with the number of channel states and the threshold levels. Simulation results show that the controller reduces joint threshold violations by \(37.3\%\) relative to adaptive AoI-only~control under the same transmission budget, while increasing mean AoI by \(8.3\%\).
\end{abstract}
\vspace{0.4em}


%
\IEEEpeerreviewmaketitle

\section{Introduction}
\label{sec_1}
\fontdimen2\font=0.64ex
The machine-type communications (MTC) landscape has evolved from delay-tolerant telemetry to large-scale sensing and real-time applications such as industrial automation, connected transportation, remote health care, and immersive services~\cite{Chettri2020}. These applications rely on sensing
devices that generate time-stamped status updates for monitoring, inference,
and control at a receiver. This shift calls for performance metrics that
capture not only packet-delivery reliability and delay but also the usefulness
of received information at decision time. One such destination-centric metric
is the age of information (AoI), which measures the time elapsed since the most
recently received update was generated~\cite{Yates2021}.

Although AoI measures information freshness, it describes only the latest
received update. Certain real-time applications also depend on the temporal
spacing between consecutive~observations. For example, a vehicle monitor may
estimate~velocity or direction from the two most recent position reports.~Even when the newest report is fresh, an older preceding report~may fail to
reflect recent acceleration, braking, or turning. AoI~cannot distinguish these
situations since it contains no information about the preceding update. Age
dispersion addresses this~limitation by measuring the separation between the
generation times of the two most recently received updates~\cite{Moltafet2026}.
Together, AoI and age dispersion indicate whether the receiver has both~a fresh
observation and two observations generated sufficiently close together.

Prior work has extensively studied update generation, packet management, and
transmission scheduling to reduce AoI over wireless
links~\cite{Sun2016,Hsu2017,Kadota2018}. Under resource constraints, studies
have considered opportunistic scheduling over multi-state time-varying channels
with power limits~\cite{Tang2020}, transmission control with automatic repeat
request (ARQ) and hybrid ARQ~\cite{Ceran2018},~and AoI-dependent packet
preemption~\cite{Banerjee2024}. Finite-state Markov models are commonly used
to capture temporal channel correlation~\cite{Sadeghi2008}. Related work has
examined AoI scheduling without current channel-state
information~\cite{Talak2018}, energy-constrained policies over time-correlated
fading~\cite{Yao2023}, and channel memory effects over a Gilbert-Elliott
erasure channel~\cite{Guan2023}. However, these efforts focus on the latest
received update rather than the temporal separation between consecutive
received updates.

Age dispersion introduces a distinct analytical and control problem. The authors of~\cite{Moltafet2026} first studied this metric in~an M/G/1/1 queue with probabilistic preemption and related~it~to higher-order AoI. Under the regenerative assumptions of~that model, the mean age dispersion reduces to the reciprocal of the delivered update throughput. This relationship does not~generally hold when channel memory creates dependence between consecutive inter-delivery intervals. Wireless channels with~the same average delivery throughput can then yield different age dispersion and joint freshness-dispersion performance. Other studies have examined joint AoI distributions~\cite{AbdElmagid2023}, the trade-off between the mean and variance of inter-delivery times~\cite{Singh2015}, and wireless scheduling policies to maintain freshness and synchronization across multiple information flows~\cite{Joo2018}. None of these studies, however, jointly address AoI and age dispersion under budget-constrained, channel-aware transmission control.

This paper investigates joint AoI and age-dispersion control for a generate-at-will link over a finite-state Markov wireless channel. For channel-aware transmission policies, we derive exact matrix expressions for the stationary joint distribution of AoI and age dispersion. We show how channel memory affects mean dispersion and joint threshold violations.~For~reversible channels, we prove that mean dispersion is at least the reciprocal of the delivery throughput and identify when equality holds. We then formulate the joint control problem as a constrained Markov decision process, derive an exact finite-state Markov chain (FSMC) model, and obtain an optimal~stationary policy through linear programming (LP) under an average transmission rate limit. Simulation results validate our analysis and demonstrate the trade-off between AoI and age dispersion over different channel memory levels and transmission budgets.

\begin{figure}[!t]
    \centering
    \includegraphics[width=\linewidth]{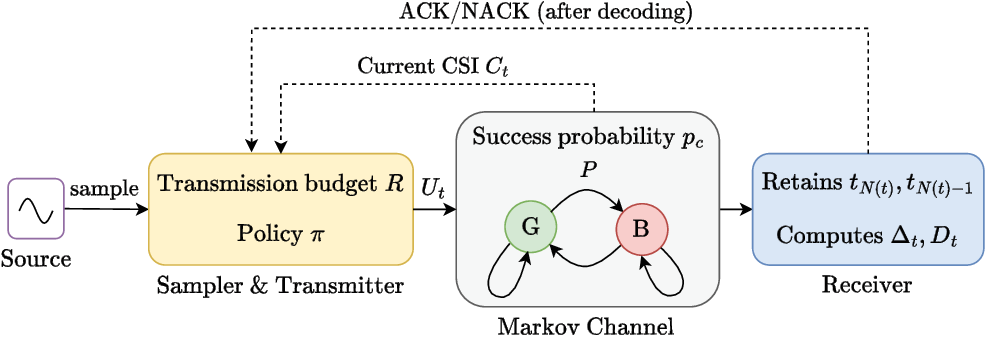}
    \vspace{-1.6em}
    \caption{Slotted status-update system over a finite-state Markov channel with CSI and \textsc{Ack}/\textsc{Nack} feedback.}
    \label{fig:system}
    \vspace{0.2em}
\end{figure}
\section{System Model and Objective}
\label{sec_2}
\fontdimen2\font=0.64ex
We consider the real-time monitoring system depicted in \figurename~\ref{fig:system}, where a single sensor observes a physical process~and sends time-stamped status updates to a receiver over a wireless link modeled as a FSMC. Time is divided into unit-length~slots $[t,t+1)$. The channel state information (CSI), denoted as $C_t\in\mathcal{C}$, evolves according to an irreducible and aperiodic Markov chain with $n=|\mathcal{C}|$ states, transition probabilities $P_{cc'}=\Pr\{C_{t+1}=c'\mid C_t=c\}$, and stationary row vector $\boldsymbol{\mu}$ satisfying $\boldsymbol{\mu}P=\boldsymbol{\mu}$. These dynamics capture temporal~correlation in packet-level channel quality.

\begin{figure}[!t]
    \centering
    \includegraphics[width=\linewidth]{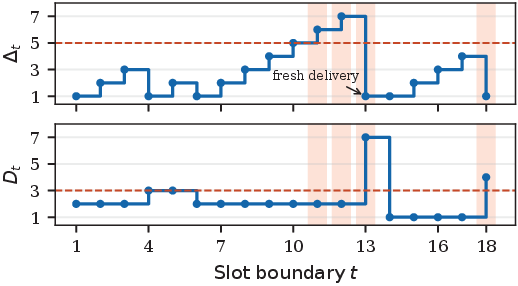}
    \vspace{-1.9em}
    \caption{Sample paths of AoI and age dispersion. Successful transmissions in slots $\{0,3,5,12,13,17\}$ are received at boundaries $\{1,4,6,13,14,18\}$, respectively. The initial retained timestamp is $-2$. Dashed lines indicate $(a_0,d_0)=(5,3)$, and shaded boundaries violate either threshold.}
    \label{fig:example}
    \vspace{0.2em}
\end{figure}

At the beginning of slot $t$, the sensor observes $C_t$ and selects $U_t\in\{0,1\}$, where $U_t=1$ denotes a transmission attempt. Under this generate-at-will operation, each attempt carries a fresh update with generation timestamp $t$. A transmission~in~channel state $c$ succeeds with probability $p_c\in(0,1)$ and, if successful, is received at $t+1$. Failed updates are discarded, and no packets are queued or retransmitted. The channel evolves~independently of the transmission decisions, while decoding trials are independent across slots conditional on the channel state sequence. Reliable \textsc{Ack}/\textsc{Nack} feedback is available before the next decision, thus allowing the sensor to reconstruct~the timestamps retained at the receiver. Each~attempt incurs one unit of transmission cost, excluding channel sensing and feedback costs. Let $t_i$ and $t_i'=t_i+1$ denote the generation and reception times of the $i$-th successfully delivered update, respectively, and let $N(t)\triangleq\max\{i\mid t_i'\leq t\}$. After two delivered updates, the AoI ($\Delta_t$) and age dispersion ($D_t$) in time slot $t$ are:
\begin{equation}
    \Delta_t=t-t_{N(t)},\qquad
    D_t=t_{N(t)}-t_{N(t)-1}.
    \label{eq:definitions}
\end{equation}
Thus, $\Delta_t+D_t$ is the age of the second latest update. Since every successful transmission occupies one time slot, the inter-delivery interval $Y_i=t_i'-t_{i-1}'=t_i-t_{i-1}$. Therefore, at each boundary, the age dynamics are:
\begin{equation}
    (\Delta_{t+1},D_{t+1})=
    \begin{cases}
        (1,\Delta_t), & \text{transmission in slot }t,\\
        (\Delta_t+1,D_t), & \text{otherwise}.
    \end{cases}
    \label{eq:reset}
\end{equation}
During the $Y_{i+1}$ slot boundaries following delivery $i$, dispersion equals $Y_i$, while AoI takes the values $\{1,\ldots,Y_{i+1}\}$. \figurename~\ref{fig:example} illustrates why a successful delivery restores freshness, but may leave a large gap between consecutive timestamps.

The receiver requires both $\Delta_t\leq a_0$ and $D_t\leq d_0$, where $a_0,d_0\geq1$ are prescribed integer thresholds. We accordingly define the violation cost $g(a,d)\triangleq\mathbf{1}\{a>a_0\text{ or }d>d_0\}$, where $\mathbf{1}\{\cdot\}$ is the indicator function.
A causal transmission policy $\pi$ may use all information available at the sensor. Its long-run violation cost and average transmission rate are:
\begin{equation}
    \begin{aligned}
        J_\pi
        &=\limsup_{T\to\infty}\frac{1}{T}
          \sum_{t=0}^{T-1}\mathbb{E}_\pi[g(\Delta_t,D_t)],\\
        \overline{U}_\pi
        &=\limsup_{T\to\infty}\frac{1}{T}
          \sum_{t=0}^{T-1}\mathbb{E}_\pi[U_t].
    \end{aligned}
    \label{eq:averages}
\end{equation}
For an average transmission rate limit $R\in(0,1]$, the control objective is expressed as:
\begin{equation}
    J^*(R)=\inf_{\pi\,\mid\,\overline{U}_\pi\leq R}J_\pi.
    \label{eq:objective}
\end{equation}
Under a stationary ergodic policy, $J_\pi=\Pr_\pi\{\Delta>a_0$ or $D>d_0\}$. The objective therefore minimizes the fraction of slot boundaries at which either the freshness or dispersion~requirement is violated, subject to the average transmission rate constraint.

\section{Joint Distribution under Channel Memory} 
\label{sec_3} 
\fontdimen2\font=0.64ex
We first analyze stationary channel-dependent randomized transmission policies. When $C_t=c$, the sensor attempts to transmit with probability $u_c\in[0,1]$, independently of previous transmission decisions and decoding outcomes. The policy therefore depends only on the current CSI. Let $h_c=u_cp_c$ denote the probability of a successful delivery in state $c$, and define $\bm h\triangleq(h_c)_{c\in\mathcal C}$. Assuming that $\bm h\ne\bm 0$, we introduce:
\begin{equation} 
    Q=P\diag(\bmone-\bm h),\quad W=P\diag(\bm h),\quad Z=(I-Q)^{-1}. 
    \label{eq:matrices} 
\end{equation}
The entry $Q_{cc'}$ is the probability that the channel state changes from $c$ to $c'$ and no delivery occurs in the next slot, whereas $W_{cc'}$ gives the corresponding probability of a successful delivery. Since $P$ is irreducible and at least one~component of $\bm h$ is positive, a delivery eventually occurs with probability one. Hence, the spectral radius of $Q$ satisfies $\operatorname{spr}(Q)<1$.

Conditioned on a successful transmission in channel state~$c$, the probability that the next delivery occurs exactly $y \geq 1$ slots later in state $c'$ is the $(c,c')$-th entry of $K_y=Q^{y-1}W$.~The channel states observed at successive delivery slots thus form an embedded Markov chain with transition matrix $G=\sum_{y\geq1}K_y=ZW$. Now, let $r=\bmu\bm h$, $\bnu=\bmu\diag(\bm h)/r$, $\bm m=Z\bmone$, and $\bar y=\bnu\bm m=1/r$. Here, $r$ is the delivery throughput, $\bnu$ is the stationary channel distribution at delivery slots, and $m_c$ is the expected time until the next delivery given that the current delivery occurs in state $c$, and $\bar y$ is the mean inter-delivery time. For integers $a,d\geq1$, we define $K_{\leq d}\triangleq\sum_{y=1}^{d}K_y=(I-Q^d)ZW$ and $\bm m_a\triangleq\sum_{k=0}^{a-1}Q^k\bmone=(I-Q^a)Z\bmone$. It can be shown that the vector $\bm m_a$ essentially gives the expected value of $\min\{Y,a\}$ conditioned on the~channel state at the preceding delivery.
\begin{theorem}[Stationary joint distribution] 
    \label{thm:law} 
    Under a stationary channel-dependent randomized transmission policy, the joint cumulative distribution function of AoI and age dispersion at an arbitrary stationary slot boundary is:
    \begin{equation} 
        \Pr\{\Delta\leq a,D\leq d\} =\frac{\bnu K_{\leq d}\bm m_a}{\bar y}. 
        \label{eq:jointcdf} 
    \end{equation} 
    The corresponding joint probability mass function is:
    \begin{equation} 
        \Pr\{\Delta=a,D=d\} =\frac{\bnu Q^{d-1}WQ^{a-1}\bmone}{\bar y}, \qquad a,d\geq1. 
        \label{eq:jointpmf} 
    \end{equation} 
\end{theorem}
\begin{proof} Using $\bmu P=\bmu$, we obtain $\bmu(I-Q)=\bmu\diag(\bm h)$. It follows that $\bnu G=\bnu$ and $\bnu\bm m =\frac{\bmu(I-Q)Z\bmone}{r} =\frac{1}{r}$. Hence, $\bnu$ is stationary for the delivery-epoch chain and the mean inter-delivery time is $\bar y=1/r$. 

Now, consider the delivery cycle beginning with update~$i$. During this cycle, $D=Y_i$, while AoI takes the values $\{1,\ldots,Y_{i+1}\}$. The number of slot boundaries satisfying both $\Delta\leq a$ and $D\leq d$ is therefore $\mathbf{1}\{Y_i\leq d\}\min\{Y_{i+1},a\}$. Starting from the channel state at delivery $i-1$, the matrix $K_{\leq d}$ accounts for $Y_i\leq d$ and the channel state at delivery $i$, while $\bm m_a$ gives the expected number of qualifying boundaries in the following cycle. The expected reward is thus $\bnu K_{\leq d}\bm m_a$. Dividing this quantity by the mean cycle length $\bar y$ proves \eqref{eq:jointcdf}. Taking finite differences with respect to $a$ and $d$ gives \eqref{eq:jointpmf}. \end{proof}

The marginal distributions follow from \eqref{eq:jointcdf} by allowing the other threshold to approach infinity. The AoI accumulated~during an inter-delivery interval of length $Y$ is $Y(Y+1)/2$. Using $\sum_{k\geq0}(k+1)Q^k=Z^2$ yields:
\begin{equation}
    \bar\Delta =\frac{\bnu Z^2\bmone}{\bar y}. 
    \label{eq:meanage}
\end{equation}
Age dispersion remains equal to $Y_i$ throughout the following interval of length $Y_{i+1}$. Its cycle reward is therefore $Y_iY_{i+1}$. Since $\sum_{y\geq1}yK_y=Z^2W$, we get:
\begin{equation}
    \bar D =\frac{\bnu Z^2W\bm m}{\bar y} =\bar y+\frac{\Cov(Y_i,Y_{i+1})}{\bar y}. 
    \label{eq:meandisp}
\end{equation}
Note that the renewal-reward argument gives the first expression in \eqref{eq:meandisp}, whereas the second follows because both inter-delivery intervals have mean $\bar y$.

When $P=\bmone\bmu$, the channel states are independent across slots and successful deliveries form a Bernoulli process with probability $r$. In this case, $\Delta$ and $D$ are independent geometric random variables on $\{1,2,\ldots\}$, and \begin{equation} \Pr\{\Delta\leq a,D\leq d\} =\bigl[1-(1-r)^a\bigr]\bigl[1-(1-r)^d\bigr]. \label{eq:iid} \end{equation} With channel memory, this product form need not hold because consecutive inter-delivery intervals can become dependent.

\begin{theorem}[Dispersion bound for reversible channels] 
    \label{thm:memory} 
    Suppose the Markov channel is reversible, such that $\mu_iP_{ij}=\mu_jP_{ji}$ for every pair of channel states. Every channel-dependent randomized policy with $r>0$ then satisfies:
    \begin{equation} 
        \bar D =\frac{1}{r}+\frac{\Var_{\bnu}(m_C)}{\bar y} \geq\frac{1}{r}, 
        \label{eq:penalty} 
    \end{equation} 
    where $C$ has distribution $\bnu$. Equality holds if and only if $m_c$ is constant over all states for which $\nu_c>0$.
\end{theorem}
\begin{proof} 
    Let $C$ denote the channel state at the successful slot separating $Y_i$ and $Y_{i+1}$. Conditioned on $C$, the channel paths before and after this slot are independent. Reversibility of the Markov channel and the conditionally independent delivery outcomes imply that the marked channel process is also reversible. The backward and forward inter-delivery intervals therefore have the same conditional mean $m_C$. Hence, $\mathbb{E}[Y_iY_{i+1}] =\mathbb{E}_{\bnu}[m_C^2]$, and  $\Cov(Y_i,Y_{i+1}) =\Var_{\bnu}(m_C)$.~Substituting them into \eqref{eq:meandisp} proves \eqref{eq:penalty}. 
\end{proof}

Theorem \ref{thm:memory} shows that delivery throughput alone does not generally determine mean dispersion. The difference between $\bar D$ and $1/r$ depends on how the expected time to the next delivery varies across the channel states observed at delivery slots. Equality can still hold in a channel with memory, including when $h_c$ is constant across states or when delivery is possible in only one state. For a non-reversible channel, consecutive inter-delivery intervals may have negative covariance, and the bound in \eqref{eq:penalty} need not hold. The joint distribution analysis also extends to the generation time separation between~the~latest update and the $k$-th preceding update. Defining $D_t^{(k)}\triangleq t_{N(t)}-t_{N(t)-k}$, we get:
\begin{equation} 
    \Pr\{\Delta\leq a,D^{(k)}\leq d\} = \frac{\bnu\!\!\!\!\!\!\!\! \displaystyle\sum_{\substack{y_1+\cdots+y_k\leq d\\ y_j \geq 1}}\!\!\!\!\!\! K_{y_1}\cdots K_{y_k}\bm m_a}{\bar y}.
    \label{eq:higher} 
\end{equation} 
In what follows, the control problem~focuses on $k=1$.

\section{Exact Finite-State Transmission Control} 
\label{sec_4} 
\fontdimen2\font=0.64ex

\subsection{Exact State Reduction}
Under adaptive transmission, the state $(\Delta_t,D_t,C_t)$ has a countably infinite state space. The objective in \eqref{eq:objective}, however, depends only on whether AoI and age dispersion exceed their respective thresholds. We define $H=\max\{a_0,d_0\}+1$, $z_t=\min\{\Delta_t,H\}$, $b_t=\mathbf{1}\{D_t>d_0\}$, and $S_t=(z_t,b_t,C_t)$. Here, $z_t$ is the capped AoI and $b_t$ indicates whether the dispersion threshold is violated. The resulting state space contains at most $2nH$ states.

For $s=(z,b,c)$, let $f_0(s,c')\triangleq(\min\{z+1,H\},b,c')$ and $f_1(s,c')\triangleq(1,\mathbf{1}\{z>d_0\},c')$ be the next states without and with a successful delivery, respectively. The transition~probability under action $u\in\{0,1\}$ is given by:
\begin{align}
    T_u(s,s')={}    &   \sum_{c'\in\mathcal C}P_{cc'}\big[(1-up_c)\mathbf{1}\{s'=f_0(s,c')\} \notag\\
                    &   \quad +up_c\mathbf{1}\{s'=f_1(s,c')\}\big].
    \label{eq:transition}
\end{align}
If no delivery occurs, AoI increases by $\min\{z+1,H\}$ and $b$ remains unchanged. Upon a delivery, \eqref{eq:reset} gives $\Delta_{t+1}=1$ and $D_{t+1}=\Delta_t$. Since $H>a_0$ and $H>d_0$, the capped~value $z_t$ preserves both threshold comparisons, and the new dispersion indicator is exactly $\mathbf{1}\{z_t>d_0\}$. As a result, the violation cost can be written as $g(s)=\mathbf{1}\{z>a_0\text{ or }b=1\}$, while the transmission cost remains $u$.

Thus, all original states represented by the same $(z,b,c)$ have identical transition probabilities and costs under either~action. The reduction is exact for the threshold objective and does not approximate \eqref{eq:objective}. The uncapped age variables are used only to evaluate mean AoI and mean age dispersion.

\subsection{Optimal Constrained Policy}
Let $x(s,u)$ denote the fraction of slots in which the stationary system is in state $s$ and action $u$ is selected. These~state-action occupation measures are obtained from~\cite{Altman2021}:
\begin{subequations}
\label{eq:lp}
\begin{align}
    \min_{x\geq0}\quad  &   \sum_{s,u}x(s,u)g(s) \label{eq:lp_obj} \\
    \textnormal{s.t.}\quad  &   \sum_u x(s',u) = \sum_{s,u}x(s,u)T_u(s,s'), \quad \forall s' \label{eq:lp_flow} \\
                            &   \sum_{s,u}x(s,u)=1 \label{eq:lp_norm} \\
                            &   \sum_s x(s,1)\leq R \label{eq:lp_budget}
    \end{align}
\end{subequations}
where constraint~\eqref{eq:lp_flow} enforces stationary flow balance, while constraint~\eqref{eq:lp_budget} limits the average transmission rate. The exact finite-state representation leading to \eqref{eq:lp} follows from the joint threshold structure of the present model.
\begin{theorem}[Attainable optimal controller] 
    \label{thm:lp} 
    Suppose that $P_{cc'}>0$ for every $c,c'$, $p_c\in(0,1)$, and $R>0$. The LP in \eqref{eq:lp} attains $J^*(R)$. For an optimal solution $x^*$, an optimal stationary policy is:
    \begin{equation} 
        \pi^*(1\mid s) =\frac{x^*(s,1)} {x^*(s,0)+x^*(s,1)} \label{eq:policy} 
    \end{equation} 
    for every occupied state. The policy transmits in unoccupied states. Moreover, an optimal basic solution can be selected such that at most one recurrent state randomizes between the two actions.
\end{theorem} 
\begin{proof} 
    Any limit point of the expected empirical occupation measures of an admissible policy satisfies \eqref{eq:lp_flow}-\eqref{eq:lp_budget}, so~the LP lower bounds~\eqref{eq:objective}. A policy attempting independently~with probability $R$ is feasible and gives $J_\pi < 1$. Thus, an optimal recurrent class contains a non-violating state. Since $P_{cc'}>0$ and every attempted transmission can fail, successive slots without delivery connect this class to every $(H,0,c)$. Recurrence, a subsequent delivery, and further failures also connect it to every $(H,1,c)$. Every closed class contains a saturated state of one of these forms, so the reconstructed policy has a unique recurrent class and realizes $x^*$.

    If $k$ states have positive occupation, their balance equations have rank $k-1$. Constraints~\eqref{eq:lp_norm} and \eqref{eq:lp_budget} permit at most $k+1$ positive variables at a basic optimum. Since every occupied state requires one positive action variable, at most one state~can use both actions.
\end{proof}


\subsection{Uncapped Mean Evaluation and Adaptive Benchmarks}
For a fixed policy $\pi(s)$, let $F$ and $L$ contain the no-delivery and delivery terms of \eqref{eq:transition}, respectively, and let $\bm\eta$ denote the stationary row vector of $F+L$. If $\operatorname{spr}(F)<1$, then define $\bm v\triangleq\bm\eta(I-F)^{-1}$ and $\bm w=\bm vL(I-F)^{-1}$.~Following~from~the stationary recursions $\bm v=\bm vF+\bm\eta$ and~$\bm w=\bm wF \!+  \bm \!vL$,~the uncapped mean AoI, mean age dispersion, and delivery throughput are, respectively, derived to be:
\begin{equation} 
    \bar\Delta=\bm v\bmone,\quad \bar D=\bm w\bmone,\quad r_\pi=\bm\eta L\bmone. 
    \label{eq:matrices} 
\end{equation}

The AoI-only controller minimizes $\Pr\{\Delta>a_0\}$ under~the same transmission budget, CSI, and feedback, using state $(\min\{\Delta,a_0+1\},C)$. Its objective is the AoI violation probability, and not mean AoI. Since several policies may attain the same minimum value $v_A^*$, their dispersion performance~can~differ. Hence, we also consider an AoI-first benchmark that~minimizes~\eqref{eq:lp} with the additional equality:
\begin{equation}
    \sum_{s,u}x(s,u)\mathbf{1}\{z_s>a_0\}=v_A^*. 
    \label{eq:add}
\end{equation}
This benchmark may use the dispersion indicator only to select among AoI-optimal policies. It gives the lowest joint violation probability compatible with the minimum AoI violation~probability and prevents the comparison from depending on an~arbitrary LP solution.
The throughput-only  policy~maximizes~the long-run successful delivery rate under the same average~transmission rate constraint, without explicitly accounting for AoI or age dispersion.

\begin{figure}[!t]
    \centering
    \includegraphics[width=0.96\columnwidth]{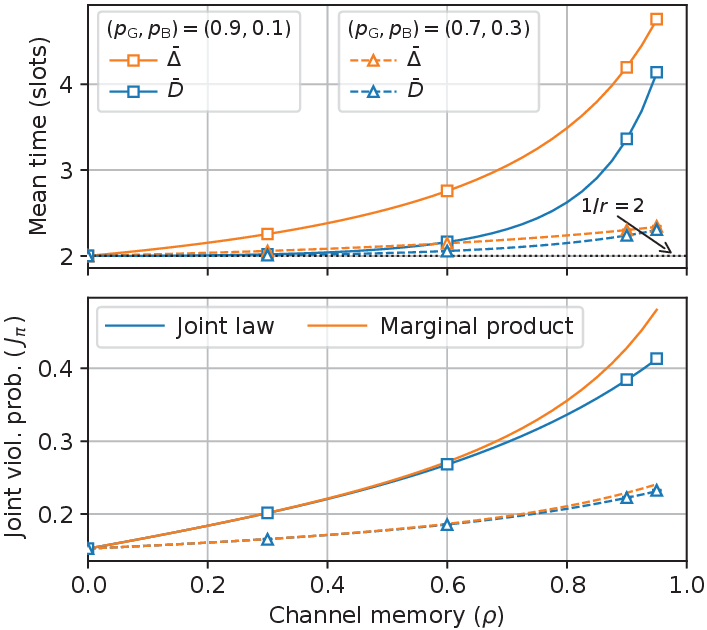}
    \vspace{-0.7em}
    \caption{Effect of channel memory at fixed throughput $r=0.5$. The lower panel uses $(a_0,d_0)=(5,3)$.}
    \vspace{-0.2em}
    \label{fig:fig3}
\end{figure}
\section{Simulation Results}
\label{sec_5}
\fontdimen2\font=0.62ex
We consider a symmetric two-state Markov channel with $P_\rho=\rho I+(1-\rho)\bmone\bmu$, where $\bmu=(1/2,1/2)$ and $\rho$ is the~lag-one correlation of the channel state indicator ($0\leq\rho<1$). Hence, $P_{\text{GG}}=P_{\text{BB}}=(1+\rho)/2$. Unless otherwise stated,~we use \(\rho=0.9\), \((p_{\text{G}},p_{\text{B}})=(0.95,0.20)\), \((a_0,d_0)=(10,6)\),~and \(B=0.4\), where \(\text{G}\) and \(\text{B}\) denote the good and bad channel states, \(p_{\text{G}}\) and \(p_{\text{B}}\) are their respective transmission success~probabilities, and \(R\) is the average transmission rate per slot. All policies observe the same current CSI and satisfy the~same transmission rate constraint. Monte Carlo (MC) results are averaged over $12$ independent runs, each containing \(10^6\) measured slots after \(5\times10^4\) warm-up slots. In all the plots, lines and markers show analytical and MC results with $95\%$ confidence intervals, respectively.

\figurename~\ref{fig:fig3} isolates the effect of channel memory over a two-state Gilbert-Elliott channel by letting the sensor transmit in every slot for $(p_{\mathrm{G}},p_{\mathrm{B}})=(0.9,0.1)$ and $(0.7,0.3)$. Both cases have $r=0.5$ and $\bar y=2$, but their good-to-bad reliability~differences are $0.8$ and $0.4$, respectively. At $\rho=0$, both yield $\bar{\Delta}=\bar D=2$ and $J_\pi=0.15$. At $\rho=0.9$, the corresponding $(\bar{\Delta},\bar D)$ values become $(4.20,3.36)$ and $(2.30,2.24)$. Therefore, a~larger reliability difference creates longer delivery gaps during persistent bad channel periods, thereby increasing the memory term $\bar D-1/r$ in Theorem~\ref{thm:memory} from $0.24$ to $1.36$. The probability~of satisfying both thresholds also falls from $84.8\%$ to $61.6\%$ and $77.8\%$, respectively. Moreover, the product-of-marginals approximation overestimates $J_\pi$ by $11.5\%$ and $3.1\%$, showing that throughput and marginal distributions cannot characterize joint performance under channel memory.

\begin{table}[!t]
\caption{Policy comparison under $R=0.4$ and $(a_0,d_0)=(10,6)$}\vspace{-0.2em}
\label{tab}
\centering
\footnotesize
\setlength{\tabcolsep}{3.3pt}
\begin{tabular}{lcccc}
\toprule
\textbf{Policy} & \textbf{$J_\pi$} & \textbf{$\Pr\{\Delta>a_0\}$} & \textbf{$\bar{\Delta}$} & \textbf{$\bar D$} \\
\midrule
Uniform random       & 0.365 & 0.175 &  6.096 & 5.298 \\
Throughput-only      & 0.363 & 0.328 & 11.632 & 2.637 \\
AoI-only             & 0.282 & 0.05 &  4.218 & 6.321 \\
AoI-first            & 0.276 & 0.05 &  4.218 & 6.288 \\
Joint controller     & 0.177 & 0.079 &  4.57 & 5.285 \\
\bottomrule
\end{tabular}
\end{table}
Table~\ref{tab} compares all policies under the same average transmission rate. The joint controller achieves \(J_\pi=0.177\), satisfying both requirements in \(82.3\%\) of the slots, compared with \(71.8\%\) and \(72.4\%\) under AoI-only and AoI-first control. This reduces \(J_\pi\) by \(37.3\%\) and approximately \(36\%\), respectively. More importantly, the dispersion-only violation probability, \(J_\pi-\Pr\{\Delta>a_0\}\), decreases from \(0.232\) under AoI-only control to \(0.098\), a \(57.8\%\) reduction. AoI-first offers only a small gain because dispersion is used merely to break AoI-optimal ties. The joint controller accepts an \(8.3\%\) increase in mean AoI to balance both objectives. In contrast, throughput-only scheduling yields the smallest mean dispersion but produces \(\bar{\Delta}=11.632\) and \(J_\pi=0.363\), confirming that neither metric alone captures the joint requirement.

\begin{figure}[t]
    \centering
    \includegraphics[width=0.95\columnwidth]{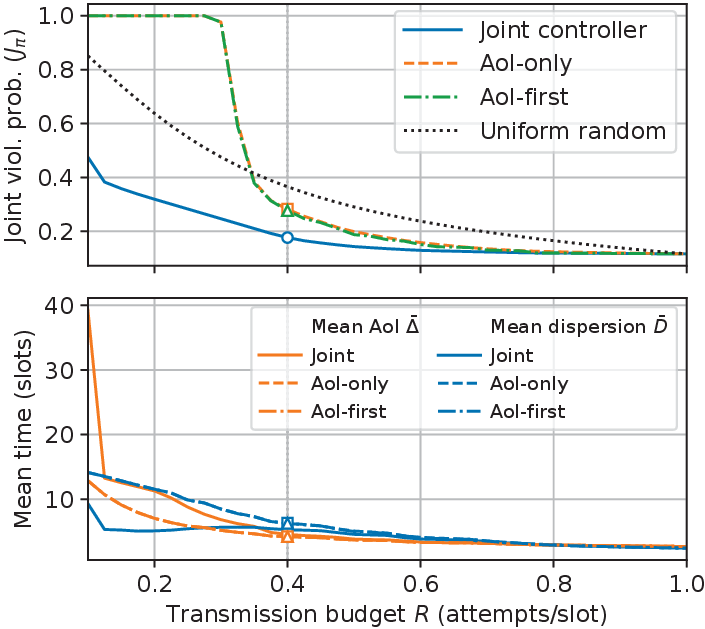}
    \vspace{-0.5em}
    \caption{Joint violation probability and uncapped mean AoI versus the~transmission budget for \(\rho=0.9\) and \((p_{\text{G}},p_{\text{B}})=(0.95,0.2)\).}
\label{fig:budget}
\end{figure}
\figurename~\ref{fig:budget} shows that transmission constraints create a clear trade-off between freshness and consecutive sample spacing. For \(R\leq 0.275\), AoI-only and AoI-first control violate the~dispersion threshold with probability one, despite maintaining lower mean AoI. At \(R=0.2\), the joint controller reduces mean dispersion from \(11.57\) to \(5.12\) slots and lowers \(J_\pi\) from \(1\) to \(0.319\), while increasing mean AoI from \(7.06\) to \(11.30\) slots. This occurs because maintaining a small dispersion requires follow-up transmissions before the current AoI becomes large, thus leaving fewer attempts for later
freshness recovery. At \(R=0.4\), the joint controller reduces \(J_\pi\) by \(37.3\%\) and mean dispersion by \(16.4\%\) relative to AoI-only control, at an \(8.3\%\) increase~in mean AoI. The policies converge as the budget~increases, showing that explicit joint control is most valuable when~transmission opportunities are limited.~Evidently, improved joint reliability need not imply lower mean AoI.

\begin{figure}[t]
    \centering
    \includegraphics[width=0.95\columnwidth]{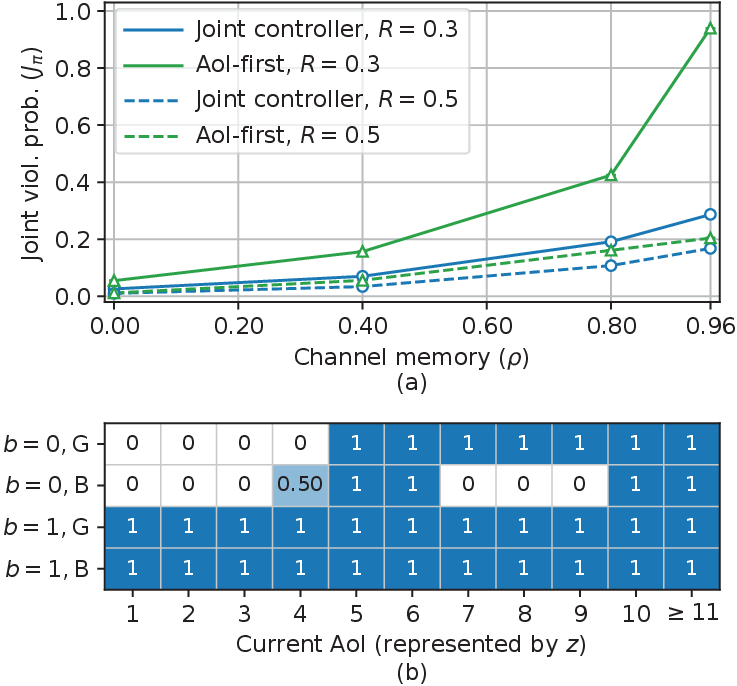}
    \vspace{-0.4em}
    \caption{Controlled performance and policy structure. (a) Joint violation probability versus channel memory for \(R\in\{0.3,0.5\}\), \((p_{\mathrm{G}},p_{\mathrm{B}})=(0.95,\,0.2)\), and \((a_0,d_0)=(10,6)\); (b) Optimal joint controller transmission probabilities at \(\rho=0.9\) and \(R=0.4\), where \(b\) is the dispersion violation indicator and \(z=\min\{\Delta,11\}\).}
\label{fig:controlled}
\end{figure}
\figurename~\ref{fig:controlled}(a) shows that dispersion-aware control becomes most valuable under strong channel memory and a tight transmission budget. For \(R=0.3\), increasing \(\rho\) from \(0\) to \(0.96\) raises~\(J_\pi\) from \(0.026\) to \(0.287\) under joint control, but from \(0.055\) to \(0.940\) under AoI-first control, yielding a \(69.5\%\) reduction~at \(\rho=0.96\). Although AoI-first has fewer AoI violations~at this point, its dispersion violation probability is \(0.929\), compared with \(0.220\) under joint control. Increasing the budget to \(R=0.5\) reduces the corresponding joint violations to \(0.168\) and \(0.204\), indicating that additional attempts partly~mitigate channel memory. \figurename~\ref{fig:controlled}(b) explains this gain. When \(b=1\), the controller always transmits to obtain the closely spaced deliveries needed to restore dispersion. When \(b=0\), it~transmits from \(z=5\) in the good state, whereas in the bad state it transmits at \(z=\{5,6\}\), pauses at \(z=\{7,8,9\}\), and resumes at~\(z\geq10\). A success during \(z=\{7,8,9\}\) would set the new dispersion to \(z>d_0\) while AoI remains acceptable. This non-monotone channel-dependent behavior arises specifically from optimizing the joint objective.

\section{Conclusion}
\label{sec_6}
\fontdimen2\font=0.66ex
In this paper, we derived the stationary joint distribution~of AoI and age dispersion over FSMC wireless channels under channel-dependent randomized transmission. Within this policy class, we showed that reversible channels yield a non-negative variance correction to reciprocal delivery throughput and identified when this correction vanishes. We also developed an exact finite-state controller that minimizes joint threshold violations under an average transmission rate constraint. The controller reduces joint violations by \(37.3\%\) relative to adaptive AoI-only control under the same transmission budget, while increasing mean AoI by \(8.3\%\). This trade-off highlights the importance of accounting for consecutive sample spacing alongside AoI. The present model assumes fresh one-slot~transmissions, current CSI, and reliable feedback. Future work will address delayed CSI and retransmissions through an expanded state representation.

\ifCLASSOPTIONcaptionsoff
  \newpage
\fi



%
\bibliographystyle{IEEEtran}
\bibliography{IEEEabrv, myref}

%

\begin{IEEEbiography}{Michael Shell}
Biography text here.
\end{IEEEbiography}

\begin{IEEEbiographynophoto}{John Doe}
Biography text here.
\end{IEEEbiographynophoto}


\begin{IEEEbiographynophoto}{Jane Doe}
Biography text here.
\end{IEEEbiographynophoto}




\end{document}